\documentclass[12pt]{article}
\usepackage{a4wide}
\usepackage{epsfig}
\usepackage{amsmath}

\newcommand{\half}{{\textstyle\frac{1}{2}}}

\newlength{\absize}
\catcode`@=11
\def\citer{\@ifnextchar
[{\@tempswatrue\@citexr}{\@tempswafalse\@citexr[]}}

\def\@citexr[#1]#2{\if@filesw\immediate
  \write\@auxout{\string\citation{#2}}\fi
  \def\@citea{}\@cite{\@for\@citeb:=#2\do
    {\@citea\def\@citea{--\penalty\@m}\@ifundefined
       {b@\@citeb}{{\bf ?}\@warning
       {Citation `\@citeb' on page \thepage \space undefined}}%
\hbox{\csname b@\@citeb\endcsname}}}{#1}}
\catcode`@=12

\begin{document}
  \thispagestyle{empty}
  \pagestyle{empty}
  \renewcommand{\thefootnote}{\fnsymbol{footnote}}
\newpage\normalsize
    \pagestyle{plain}
    \setlength{\baselineskip}{4ex}\par
    \setcounter{footnote}{0}
    \renewcommand{\thefootnote}{\arabic{footnote}}
\newcommand{\preprint}[1]{%
  \begin{flushright}
    \setlength{\baselineskip}{3ex} #1
  \end{flushright}}
\renewcommand{\title}[1]{%
  \begin{center}
    \LARGE #1
  \end{center}\par}
\renewcommand{\author}[1]{%
  \vspace{2ex}
  {\Large
   \begin{center}
     \setlength{\baselineskip}{3ex} #1 \par
   \end{center}}}
\renewcommand{\thanks}[1]{\footnote{#1}}
\begin{flushright}
\end{flushright}
\vskip 0.5cm

\begin{center}
{\large \bf Spectrum of $q$-Deformed Schr\"odinger Equation}
\end{center}
\vspace{1cm}
\begin{center}
Jian-zu Zhang$^{a,b,*}$
\end{center}
\vspace{1cm}
\begin{center}
$^a$ Institute for Theoretical Physics, Box 316, East China
University of Science and Technology, Shanghai 200237, P. R.
China \\
$^b$ The Abdus Salam International Centre for Theoretical Physics,
P. O. Box 586, I-34014 Trieste, Italy \\
\end{center}
\vspace{1cm}

\begin{abstract}
The energy spectrum of $q$-deformed Schr\"odinger equation is
demonstrated. This spectrum includes an exponential factor with
new quantum numbers--the $q$-exciting number and the scaling
index. The pattern of quark and lepton masses is qualitatively
explained by such a $q$-deformed spectrum in a composite model.

\end{abstract}

\begin{flushleft}
${^*}$ E-mail address: jzzhang@ecust.edu.cn\\
\end{flushleft}
\clearpage
Quarks and leptons are considered as point-like particles which is
correct at least down to $10^{-17} cm.$ The regularities found in
the quark and lepton mass spectrum and several other motivations
suggest a possibility that at short distances much smaller than
$10^{-17} cm$ quarks and leptons are composites of some basic
entities. For an earlier consideration of a substructure, see e.
g. Refs.~\cite{Pati,Harari}. One of the common problem of
composite models is to understand the mysterious mass spectrum of
quarks and leptons. The striking features of the quark-lepton mass
spectrum are that it is similar (the family structure and the
generation structure) and has a large mass range (the mass ratios
reach as large as $10^{5}$ order) with exponential interval.

Because such substructures involve distances which are many orders
of magnitude below ones of present physics, dynamics at very short
distances may be radically different, and is likely to involve
some entirely new principles. According to the present tests of
quantum electrodynamics, quantum theories based on Heisenberg's
commutation relation (Heisenberg's algebra) are correct at least
down to $10^{-17}$~cm. For possible new quantum theories it is
likely that a modification of Heisenberg's algebra must be at
short distances much smaller than $10^{-17}$~cm. In searching for
such a possibility at short distances consideration of the space
structure is a useful guide.

Recently $q$-deformed quantum mechanics is proposed
\citer{Schwenk,JZZ98} in the framework of quantum group. Quantum
groups are a generalization of symmetry groups which we have
successfully used in physics. A general feature of space carrying
a quantum group structure is that they are noncommutative and
inherit a well-defined mathematical structure from quantum group
symmetries. There is a possibility that noncommutativity of space
might be a realistic picture of space at short distances.
$q$-deformed quantum mechanics is based on the $q$-deformed
Heisenberg algebra \cite{Hebecker,FLW} which is a generalization
of Heisenberg's algebra. Starting from the $q$-deformed Heisenberg
algebra a general dynamical equation of $q$-deformed quantum
mechanics is obtained \cite{JZZ98}. A general feature of this
equation is that its energy spectrum shows an exponential
$q$-structure \citer{FLW,JZZ98}. In this letter we report that the
pattern of the quark-lepton masses can be qualitatively explained
by such a $q$-deformed spectrum in a composite model, for example,
the rishon model \cite{Harari}. The calculated quark and lepton
masses agree with known data.

In the $q$-deformed phase space Refs.~\cite{Hebecker,FLW}
generalized the Heisenberg algebra to  the following $q$-deformed
Heisenberg algebra:
\begin{equation}
\label{Eq:q-algebra} q^{1/2}XP-q^{-1/2}PX=i\Lambda^{-1}, \quad
\Lambda X=qX\Lambda, \quad \Lambda P=q^{-1}P\Lambda,
\end{equation}
where the position $X$ and the momentum $P$ are hermitian,
$\Lambda$ is unitary:
\begin{equation}
\label{Eq:hermitian} X^{\dagger}=X, \quad P^{\dagger}=P, \quad
\Lambda^{\dagger}=\Lambda^{-1}.
\end{equation}
In (\ref{Eq:q-algebra}) the parameter $q$ is real and $q>1$,
$\Lambda$ is called scaling operator. The algebra
(\ref{Eq:q-algebra}) is based on the definition of the hermitian
$P$. However, if $X$ is assumed to be a hermitian operator in a
Hilbert space, the usual quantization rule $P\to -i\partial_X$
does not yield a hermitian momentum operator. Ref.~\cite{FLW}
showed that a hermitian momentum operator $P$ is related to
$\partial_X$ and $X$ in a nonlinear way by introducing a scaling
operator $\Lambda$
\begin{equation*}
\Lambda\equiv q^{1/2}[1+(q-1)X\partial_X], \quad
\bar\partial_X\equiv -q^{-1/2}\Lambda^{-1}\partial_X, \quad
P\equiv -\frac{i}{2}(\partial_X-\bar\partial_X), \nonumber
\end{equation*}
where $\bar\partial_X$ is the conjugate of $\partial_X$. Because
the scaling operator $\Lambda$ is introduced in the definition of
the hermitian momentum operator, it closely relates to properties
of dynamics and plays an essential role in $q$-deformed quantum
mechanics. The nontrivial properties of  $\Lambda$ leads to that
the algebra (\ref{Eq:q-algebra}) has a richer structure than the
Heisenberg algebra. In the case of $q$ approaching to one the
scaling operator $\Lambda$ reduces to the unit operator, thus the
algebra (\ref{Eq:q-algebra}) reduces to the Heisenberg algebra.

The variables $X$, $P$ of the $q$-deformed algebra
(\ref{Eq:q-algebra}) can also be expressed in terms of the
variables of an undeformed algebra. There are three pairs of
canonically conjugate variables \cite{FLW}:

1. The variables $\hat x$ and $\hat p$ of the undeformed quantum
mechanics which satisfy: $[ \hat x, \hat p ]=i$.

2. The variables $\tilde x$ and $\tilde p$ which are obtained by a
canonical transformation of $\hat x$ and $\hat p$: $\tilde p=
f(\hat z)\hat p$, $\tilde x=\hat x f^{-1}(\hat z)$ where
\begin{equation}
\label{Eq:f(z)} f^{-1}(\hat z)= \frac{[\hat z-\half]}{\hat
z-\half},\quad \hat z=-\frac{i}{2}(\hat x\hat p+\hat p\hat x),
\end{equation}
and $[A]=(q^A-q^{-A})/(q-q^{-1})$. The function defined by
(\ref{Eq:f(z)}) satisfy $\hat x f(\hat z)=f(\hat z+1)\hat x$,
$\hat p f(\hat z)=f(\hat z-1)\hat p$. The variables $\tilde x$ and
$\tilde p$ also satisfy the same commutation relation as $\hat x$
and $\hat p$: $[ \tilde x, \tilde p ]=i$. Thus in the $\tilde x$
representation $\tilde p =-i\tilde\partial$, where
$\tilde\partial=\partial/\partial {\tilde x}$, and all the
machinery of quantum mechanics can be used for the ($\tilde x$,
$\tilde p$) system.

3. The $q$-deformed variables $X$ and $P$, where $X$, $P$ and the
scaling operator $\Lambda$ are related to  $\tilde  x$ and $\tilde
p$ in the following way:
\begin{equation}
\label{Eq:X-x} X=\tilde x, \quad P=f^{-1}(\tilde z) \tilde p,
\quad \Lambda= q^{-\tilde z}.
\end{equation}
In (\ref{Eq:X-x}) $\tilde z$ and  $f^{-1}(\tilde z)$ are defined
by the same equations (\ref{Eq:f(z)}) for $\hat  z$ and
$f^{-1}(\hat z)$. It is easy to check that  $X$, $P$ and $\Lambda$
in (\ref{Eq:X-x}) satisfy (\ref{Eq:q-algebra}) and
(\ref{Eq:hermitian}).

In order to derive the dynamical equation of $q$-deformed quantum
mechanics our starting point is to use the $q$-deformed variables
to write down the Hamiltonian in analogy with the undeformed
system, then using (\ref{Eq:X-x}) to represent $X$, $P$ and
$\Lambda$ by $\tilde x$ and $\tilde p$. From (\ref{Eq:q-algebra})
and (\ref{Eq:hermitian}) it follows that
\begin{equation}
\label{Eq:P2}
P^2=-(q-q^{-1})^{-2}X^{-2}\left[q^2\Lambda^{-2}-(q+q^{-1})
+q^{-2}\Lambda^2\right].
\end{equation}
Using (\ref{Eq:X-x}) and (\ref{Eq:P2}) we obtain the stationary
dynamical equation of $q$-deformed quantum mechanics
\begin{equation}
\label{Eq:eq1} \left\{-\frac{1}{2\mu}(q-q^{-1})^{-2}\tilde
x^{-2}\left[q(q^{-2\tilde x\tilde\partial}-1)+q^{-1}(q^{2\tilde
x\tilde\partial}-1)\right]+V(\tilde x)\right\}\psi(\tilde x)
=E_q\psi(\tilde x).
\end{equation}
Eq.~(\ref{Eq:eq1}) is a non-linear equation, which is a
$q$-generalization of the Schr\"odinger equation.

For the case of $q$ closing to 1, let $q=e^{f}$, $0<f\ll 1$. The
perturbation expansion of (\ref{Eq:eq1}) reads
\begin{eqnarray}
\label{Eq:H1} H
=&-&\frac{1}{2\mu}(2f+\frac{1}{3}f^3+\cdots)^{-2}\tilde
x^{-2}\bigl[4f^2\tilde x^2\tilde\partial^2
\nonumber \\
&+&\frac{1}{3}f^4(4\tilde x^4\tilde\partial^4+16\tilde
x^3\tilde\partial^3+10\tilde x^2\tilde\partial^2)+\cdots
\bigr]+V(\tilde x).
\end{eqnarray}
To the lowest order of $f$, (\ref{Eq:eq1}) reduces to the
Schr\"odinger equation of the ($\tilde x$, $\tilde p$) system
\begin{equation}
\label{Eq:eq2}\left[-\frac{1}{2\mu}\tilde\partial^2 +V(\tilde
x)\right]\psi(\tilde x) =E\psi(\tilde x).
\end{equation}
In (\ref{Eq:H1}) the next order correction of $H$ shows a complex
structure which amounts to some additional momentum dependent
interaction. As an example, we consider the harmonic oscillator
potential $V(\tilde x)=\frac{1}{2}\mu\omega^2 \tilde x^2$. The
spectrum of the un-perturbation Hamiltonian
$H_0=-\frac{1}{2\mu}\tilde\partial^2 +\frac{1}{2}\mu\omega^2
\tilde x^2$ is equal interval,
$E^{(0)}_n=\omega(n+\frac{1}{2})$,($n=0, 1, 2,\cdots $). The
perturbation Hamiltonian $H_I$ in (\ref{Eq:H1}) is
\begin{equation}
\label{Eq:qd-harmonic} H_I=-\frac{f^2}{12\mu}(2\tilde
x^2\tilde\partial^4+8\tilde x\tilde\partial^3+3\tilde\partial^2).
\end{equation}
The perturbation shifts of the energy levels are $\Delta E^{(0)}_n
=\int d\tilde x\psi^{(0)*}_n(\tilde x) H_I\psi^{(0)}_n(\tilde x)$,
where
\begin{equation*}
\psi^{(0)}_n(\tilde
x)=2^{-n/2}(n!)^{-1/2}(\mu\omega/\pi)^{1/4}exp(-\mu\omega\tilde
x^2/2)H_n(\sqrt{\mu\omega}\tilde x) \nonumber
\end{equation*}
and $H_n$ is the Hermite polynomials. The first few shifts are
$\Delta E^{(0)}_0=-3f^2\omega/16$, $\Delta
E^{(0)}_1=-5f^2\omega/12$, $\Delta E^{(0)}_2=-f^2\omega/2$, etc.
Thus the intervals of the total spectrum $E_n=E^{(0)}_n+\Delta
E^{(0)}_n$ are not equal.

For the non-perturbation properties of the $q$-deformed system, we
consider a simple case, the "free" Hamiltonian
$H_0=\frac{1}{2\mu}P^2$. Suppose that the eigenvalue of $H_0$ is
solved: $H_0|\epsilon_0\rangle=\epsilon_0|\epsilon_0\rangle$ with
the normalization condition of $|\epsilon_0\rangle$, $\langle
\epsilon^{\prime}_0 |\epsilon_0\rangle=\delta (\epsilon^{\prime}_0
-\epsilon_0)$. $H_0$ is semi-positive definite, i.e.
$\epsilon_0\ge 0$. The state $|\epsilon_0\rangle$ is a common
eigenstate of $H_0$ and the momentum $P$, $P|\epsilon_0\rangle=\pm
(2\mu \epsilon_0)^{1/2}|\epsilon_0\rangle$, here the plus and
minus sign correspond, respectively, to the right and left moving
modes. Now we consider the action of the scaling operator
$\Lambda$ on the state $|\epsilon_0\rangle$. From the algebra
(\ref{Eq:q-algebra}) we have $H_0(\Lambda^M|\epsilon_0\rangle)=
\epsilon_0q^{2M}(\Lambda^M|\epsilon_0\rangle)$. i. e.
$\Lambda^M|\epsilon_0\rangle=|\epsilon_0 q^{2M}\rangle$ with the
normalization condition $\langle \epsilon^{\prime}_0
q^{2M^{\prime}} |\epsilon_0 q^{2M}\rangle=\delta_{M^{\prime}M}
\delta (\epsilon^{\prime}_0 -\epsilon_0)$. Thus in the general
case the $q$-deformed spectrum of $H_0$ is $H_0|\epsilon_0 q^{2nM}
\rangle= E_n|\epsilon_0 q^{2nM} \rangle$, $|\epsilon_0 q^{2nM}
\rangle=(\Lambda^M)^n|\epsilon_0\rangle$, $E_n=\epsilon_0
q^{2nM}$, where $n=0, 1, 2,\cdots$; $M=0, 1, 2,\cdots$. The
undeformed energy $\epsilon_0$ is determined by dynamics, the
exponential factor $q^{2nM}$ (the $q$-exciting structure) is
determined by the algebra (\ref{Eq:q-algebra}). The non-trivial
properties of the scaling operator $\Lambda$ leads to a richer
structure of the algebra (1) than the Heisenberg algebra and, as a
result, leads to the $q$-structure of the spectrum. This spectrum
includes new quantum numbers, the $q$-exciting number $n$ and the
scaling index $M$. In the case of $q=1$, which corresponds to the
present-day physics, the $q$-exciting degree of freedom freezes,
and the $q$-exciting spectrum $E_n$ reduces to the undeformed one
$\epsilon_0$.

In the following in a composite scheme we use the $q$-deformed
spectrum to qualitatively explain the mass pattern of quarks and
leptons. The only strong gauge interaction that we understand in
any detail is $QCD$. But for the non-perturbation aspect of the
strong coupling it is not clear in $QCD$ how to calculate bound
states of quarks to yield the hadron spectrum except for some
simple cases which can be treated by lattice $QCD$. Thus it is
helpful to treat the hadron spectrum by some phenomenological
approaches which may guide us to the right direction. In analogy
with calculations of the hadron mass spectrum we calculate bound
states of substructure to yield the quark-lepton mass spectrum in
a phenomenological approach. Suppose that the composite system is
described by the Hamiltonian $H=H_0-V_0$, where $H_0$ is the
"free" Hamiltonian and $V_0>0$ is the binding energy. In order to
reduce the number of phenomenological parameters we consider a
simplified example,the rishon model \cite{Harari} in which the
most economical set of building blocks consists of two spin
$J=1/2$ rishons, a charged $T (Q=1/3)$ and a neutral $V (Q=0)$.
Their antiparticles are $\bar T (Q=-1/3)$ and $\bar V (Q=0)$. The
first generation of composite fermions is $u$-quark $(TTV, TVT,
VTT)$, $d$-quark $(\bar T\bar V\bar V, \bar V\bar T\bar V, \bar
V\bar V\bar T)$, neutrino $\nu_e(VVV)$ and electron $e(\bar T\bar
T\bar T)$. The dynamics is supposed to be that the three states of
quarks are degenerate. The $q$-deformed spectrum in the above
simplified example is
\begin{equation}
\label{Eq:H0} H|\epsilon^{(i)}_0 q^{2nM_i} \rangle=
E^{(i)}_n|\epsilon^{(i)}_0 q^{2nM_i} \rangle,
\end{equation}
\begin{equation}
\label{Eq:En} E^{(i)}_n=\epsilon^{(i)}_0 q^{2nM_i}-V^{(i)}_0,\quad
(n=0, 1, 2,\cdots; M_i=0, 1, 2,\cdots),
\end{equation}
\begin{equation}
\label{Eq:epsilon0} \epsilon^{(1)}_0=2\mu_T+\mu_V,\quad
\epsilon^{(2)}_0=\mu_T+2\mu_V, \quad \epsilon^{(3)}_0=3\mu_T,
\quad \epsilon^{(4)}_0=3\mu_V,
\end{equation}
\begin{equation}
\label{Eq:V0} V^{(1)}_0=V^{(2)}_0=V^{(3)}_0=V_0,\quad
V^{(4)}_0=V^{\prime}_0.
\end{equation}
The physical contents of the spectrum (\ref{Eq:H0})-(\ref{Eq:V0})
are as follows. The index $i=1, 2, 3, 4$ represents families, i.
e. the scaling indices $M_1$, $M_2$ and $M_3$, $M_4$ correspond
to, respectively, the quark families $(u, c, t)$, $(d, s, b)$ and
the lepton families $(e, \mu, \tau)$, $(\nu_e, \nu_{\mu},
\nu_{\tau})$. The $q$-exciting quantum numbers $n=0, 1, 2$
correspond to, respectively, the first, the second and the third
generation $(u, d; \nu_e, e)$, $(c, s; \nu_{\mu}, \mu)$ and $(t,
b; \nu_{\tau}, \tau)$
\footnote{In composite models the next generation are simply
considered as higher excitations of the first generation. At the
present stage one of the common open problems in composite models
is that there is no principle to govern a choice of the value $n$
of generations. The only way of fixing the maximum value of $n$ is
the experimental results from measurements at the $Z$ peak. These
measurements establish the number of light neutrino generations to
be $n_{\nu}=2.994\pm 0.012$ (Standard Model fits to LEP data) and
$n_{\nu}=3.07\pm 0.12$ (Direct measurement of invisible $Z$)
\cite{Gonz}.}
. The undeformed spectrum $\epsilon^{(i)}_0$ are flavor dependent,
and $\mu_T$ and $\mu_V$ are, respectively, the ground state
energies of the "free" $T$ and $V$. The binding energy
$V^{\prime}_0$ of the neutrino family is supposed to be different
from the other ones.

Eq.~(\ref{Eq:epsilon0}) gives a simple regularity among $m_u$,
$m_d$ and $m_e$
\begin{equation}
\label{Eq:m-ude} 2m_u-m_d=m_e.
\end{equation}
which we were not previously aware of.

If we put in
\begin{eqnarray}
\label{Eq:qMi} &&q^{2M_1}=128.98,\quad q^{2M_2}=17.28,\quad
q^{2M_3}=15.90,\nonumber\\
&&\mu_T=2.35\, MeV,\quad \mu_V=5.70\, MeV,\nonumber\\
&&V_0=6.55\, MeV.
\end{eqnarray}
Eqs.~(\ref{Eq:En})-(\ref{Eq:V0}) give (in $MeV$ units)
\begin{eqnarray}
\label{Eq:mi1}1.\; m_u&=&E^{(1)}_1=4 \;(1.5-5),\nonumber\\
m_c&=&E^{(1)}_2=1340 \;(1100-1400),
\nonumber\\
m_t&=&E^{(1)}_3 =170\times 10^3 \;\left({(73.8\pm 5.2)\times 10^3
 \atop (170\pm 7(+14))\times
10^3 }\right)
\end{eqnarray}
\begin{eqnarray}
\label{Eq:mi2} 2.\;m_d&=&E^{(2)}_1=7 \;(3-9),\nonumber\\
m_s&=&E^{(2)}_2=230 \;(60-170),
\nonumber\\
m_b&=&E^{(2)}_3 =4100 \;(4100-4400).
\end{eqnarray}
\begin{eqnarray}
\label{Eq:mi3}3.\; m_e&=&E^{(3)}_1=0.5 \;(0.51099907\pm 0.00000015),\nonumber\\
m_{\mu}&=&E^{(3)}_2=106 \;(105.658389\pm 0.000034),
\nonumber\\
m_{\tau}&=&E^{(3)}_3 =1777 \;\left(1777.05\qquad {+0.29\atop
-0.26}\right).
\end{eqnarray}
And
\begin{eqnarray}
\label{Eq:m-m} &&m_u/m_d=0.54 \;(0.20-0.70),\nonumber\\
&&m_s/m_d=32 \;(17-25),\nonumber\\
&&\bar m=(m_u+m_d)/2=5.6 \;(2-6),
\nonumber\\
&&[m_s-(m_u+m_d)/2]/(m_d-m_u)=68 \;(34-51).
\end{eqnarray}
In (\ref{Eq:mi1})-(\ref{Eq:m-m}) the data in the brackets are
cited from Ref.~\cite{Data}. (In (\ref{Eq:mi1}) for top quark the
datum in the first line from direct observation of top events; the
one in the second line from Standard Model electroweak fit). The
calculated masses agree with known data
\footnote{The concept of quark mass is involved. The values of the
quark masses depend on the energy scale where they are calculated.
As in the quark model of hadrons, the free parameters in
(\ref{Eq:En})-(\ref{Eq:V0})  are a phenomenological input of the
theory. In (\ref{Eq:mi1}) and (\ref{Eq:mi2}) the values of $m_u$,
$m_t$ and $m_b$ cited from Ref.~\cite{Gonz} are used to fix the
parameters. Thus the energy scale of the calculated quark masses
in (\ref{Eq:mi1}) and (\ref{Eq:mi2}) is related to the energy
scale of quark masses cited in Ref.~\cite{Gonz}.}
.

At present little about neutrino masses can be predicted because
of lack of definite data. In Standard Model of particles neutrinos
could be exactly massless, although this would violate naive
quark-lepton symmetry. According to (\ref{Eq:En}) massless
neutrinos correspond to $\epsilon^{(4)}_0=V^{(4)}_0,\;
q^{2M_4}=1$. There are several hints for non-vanishing neutrino
masses which can be inferred from the observations of the solar
neutrinos \cite{Fogli}, the atmospheric neutrinos \cite{Gonz} and
the results of LSND \cite{Atha}. These data can be understood in
terms of neutrino oscillations which depend on the different
neutrino mass-squared differences $\Delta m^2_{ij}$. The solution
of the MSW type \cite{Wolf} to the solar neutrino puzzle yields
the so-called small angle solution \cite{Fogli} $\Delta
m^2_{ei}\sim 4\times 10^{-6}-1.2\times 10^{-5}\, eV^2$. Assuming
$\nu_{\mu}-\nu_{\tau}$ oscillation the presently available
atmospheric neutrino data yields \cite{Gonz} $\Delta
m^2_{\mu\tau}\sim 4\times 10^{-4}-5\times 10^{-3}\, eV^2$.
Finally, the LSND data suggests \cite{Atha} that $\Delta
m^2_{e\mu}\sim 0.2-10\, eV^2$. The Solar MSW small angle and
atmospheric neutrino along \cite{LSND,MSW} indicate very small
differences between the neutrino masses
\footnote{For the mass-squared differences $\Delta m^2_{ei}$ in
the MSW small angle result, the type of neutrino $\nu_i$ is,
depending on the specific version of the effects, a $\nu_{\mu}$, a
$\nu_{\tau}$, a $\nu_{\mu}-\nu_{\tau}$ mixture, or perhaps a
sterile neutrino $\nu_s$.}
. We may suggest a degenerate scheme where all three masses are
large relative to their splitting and almost degenerate. There is
no clear way to set a meaningful limit on $m_{\nu_e}$. If the
three neutrinos are the candidate for the hot dark matter, an
estimation of the total mass of neutrinos is about $4.8\;eV$
\cite{Schm}. In (\ref{Eq:En})-(\ref{Eq:V0}) the estimations of
$m_{\nu_e}\sim m_{\nu_{\mu}}\sim m_{\nu_{\tau}}\sim 1.6\;eV$,
$\Delta m^2_{e\mu}\sim \Delta m^2_{\mu\tau}\sim 10^{-4}\, eV^2$
correspond to inputs $3\mu_V-V_0^{\prime}\sim 1.6\;eV$,
$q^{2M_4}-1\sim 10^{-10}$.

In (\ref{Eq:mi1})-(\ref{Eq:mi3}) 9 observed masses are explained
by a fit to 6 free parameters, which are a phenomenological input
of the theory. If we find an effective way to solve the nonlinear
$q$-deformed Schr\"odinger equation (\ref{Eq:eq1}), the parameters
$\mu_T$, $\mu_V$ and $V_0$ are expected to be calculable.

If $q$-deformed quantum mechanics is a correct theory, its effects
mainly manifest at short distances much smaller than
$10^{-17}\;cm$. At such short distances if rishon dynamics is
governed by a $q$-deformed gauge theory, we may expect a better
explanation of the quark-lepton mass spectrum by bound state
calculation in $q$-deformed gauge theory. Of course, this will be
very difficult topics, perhaps much difficult than bound state
treatment in $QCD$.

In summary, in this letter we show that the new degrees of freedom
in the $q$-deformed spectrum emerge. The qualitative explanation
of the mass spectrum of quarks and leptons by this spectrum is
encouragement which may guide us to the right direction in
understanding dynamics at very short distances.

\vspace{0.4cm}

{\bf Acknowledgements}

\vspace{0.4cm}

The author would like to thank J. Wess very much for his many
stimulating helpful discussions, N. Schmitz and G. Raffelt for
helpful discussions on neutrino masses. He would like to thank M.
A. Virasoro and S. Randjbar-Daemi for hospitality at the Abdus
Salam International Centre for Theoretical Physics. His work has
also been supported by the National Natural Science Foundation of
China under the grant number 19674014 and by the Shanghai
Education Development Foundation.

\vspace{0.4cm}

\end{document}